\input lanlmac
\input epsf.tex
\input mssymb.tex
\input rs-hs.defs
\font\sc=cmcsc10
\overfullrule=0pt
\def\figbox#1#2{\epsfxsize=#1\vcenter{%
\hbox{\epsfbox{#2}}}}
\newcount\figno
\figno=1
\def\fig#1#2#3{%
\xdef#1{\the\figno}%
\writedef{#1\leftbracket \the\figno}%
\nobreak%
\par\begingroup\parindent=0pt\leftskip=1cm\rightskip=1cm\parindent=0pt%
\baselineskip=11pt%
\midinsert%
\centerline{#3}%
\vskip 12pt%
{\bf Fig.\ \the\figno:} #2\par%
\endinsert\endgroup\par%
\goodbreak%
\global\advance\figno by1%
}
\newcount\tabno
\tabno=1
\def\table#1#2#3{%
\xdef#1{\the\tabno}%
\writedef{#1\leftbracket \the\tabno}%
\nobreak%
\par\begingroup\parindent=0pt\leftskip=1cm\rightskip=1cm\parindent=0pt%
\baselineskip=11pt%
\midinsert%
\centerline{#3}%
\vskip 12pt%
{\bf Tab.\ \the\tabno:} #2\par%
\endinsert\endgroup\par%
\goodbreak%
\global\advance\figno by1%
}
\long\def\rem#1{}
\def\e#1{{\rm e}^{#1}}
\def\pre#1{{\tt
#1}}
\def\tr{{\rm tr}}

\def\der{\partial}
\def\bra#1{\langle #1 |}
\def\ket#1{| #1 \rangle}
\def\braket#1#2{\langle #1 | #2 \rangle}
\def\LP{{\cal L}_{2n}}
\def\LPl{{\cal L}_{\ell,2m}}
\def\LPlo#1{{\cal L}_{\ell,#1}}
\def\H{{\cal H}_{2n}}
\def\Hl{{\cal H}_{\ell,2m}}
\def\Hlaux{{\cal H}_{\ell,2m+1}}
\def\alphal{\tilde{\alpha}}
\def\betal{\tilde{\beta}}
\newcount\propno
\propno=1
\def\prop#1#2\par{\xdef#1{\the\propno}%
\medbreak\wrlabeL#1%
\noindent{\sc Proposition \the\propno.\enspace}{\sl #2}\medbreak%
\global\advance\propno by1}
\newcount\thmno
\thmno=1
\def\thm#1#2\par{\xdef#1{\the\thmno}%
\medbreak\wrlabeL#1%
\noindent{\sc Theorem \the\thmno.\enspace}{\sl #2\par}\medbreak%
\global\advance\thmno by1}
\newcount\conjno
\conjno=1
\def\conj#1#2\par{\xdef#1{\the\conjno}%
\medbreak\wrlabeL#1%
\noindent{\sc Conjecture \the\conjno.\enspace}{\sl #2}\medbreak%
\global\advance\conjno by1}
\newcount\lemno
\lemno=1
\def\lemma#1#2\par{\xdef#1{\the\lemno}%
\medbreak\wrlabeL#1%
\noindent{\sc Lemma \the\lemno.\enspace}{\sl #2}\medbreak%
\global\advance\lemno by1}
\def\corol#1\par{%
\medbreak\noindent{\sc Corollary.\enspace}{\sl #1}\medbreak}
\def\example#1\par{%
\medbreak\noindent{\sc Example:\enspace}#1\medbreak}
\def\qed{\nobreak\hfill\vbox{\hrule height.4pt%
\hbox{\vrule width.4pt height3pt \kern3pt\vrule width.4pt}\hrule height.4pt}\medskip\goodbreak}
\lref\Kho{M.~Khovanov, 
{\sl Graphical calculus, canonical bases and Kazhdan-Lusztig theory}, 
PhD dissertation (1997).}
\lref\Kup{G. Kuperberg, {\sl Symmetry classes of alternating-sign matrices under one roof},
{\it Ann. of Math.} (2) 156 (2002), no. 3, 835--866,
\pre{math.CO/0008184}.}
\lref\Oka{S. Okada, {\sl  Enumeration of Symmetry Classes of Alternating Sign Matrices and Characters of Classical Groups}, \pre{math.CO/0408234}.}
\lref\Kit{A. Caradoc, O. Foda, N. Kitanine,
{\sl Higher spin vertex models with domain wall boundary conditions},
\pre{math-ph/0601061}.}
\lref\Lu{G. Luzstig, {\sl Canonical bases arising from quantized enveloping algebras}, {\it J. Amer. Math. Soc.} 3 (1990), 447--498.}
\lref\RS{A.V. Razumov and Yu.G. Stroganov, 
{\sl Combinatorial nature
of ground state vector of $O(1)$ loop model},
{\it Theor. Math. Phys.} 
{\bf 138} (2004) 333-337; {\it Teor. Mat. Fiz.} 138 (2004) 395-400, \pre{math.CO/0104216}.}
\lref\BdGN{M.T. Batchelor, J. de Gier and B. Nienhuis,
{\sl The quantum symmetric XXZ chain at $\Delta=-1/2$, alternating sign matrices and 
plane partitions},
{\it J. Phys.} A34 (2001) L265--L270,
\pre{cond-mat/0101385}.}
\lref\DFZJ{P.~Di Francesco and P.~Zinn-Justin, {\sl Around the Razumov--Stroganov conjecture:
proof of a multi-parameter sum rule}, {\it E. J. Combi.} 12 (1) (2005), R6,
\pre{math-ph/0410061}.}
\lref\DFZJb{P.~Di Francesco and P.~Zinn-Justin, {\sl Inhomogeneous model of crossing loops
and multidegrees of some algebraic varieties}, 
{\it Commun. Math. Phys.} 262 (2006), 459--487,
\pre{math-ph/0412031}.}
\lref\KZJ{A. Knutson and P. Zinn-Justin, {\sl A scheme related to the Brauer loop model}, \pre{math.AG/0503224}.}
\lref\Pas{V.~Pasquier, {\sl Quantum incompressibility and Razumov Stroganov type conjectures},
\pre{cond-mat/0506075}.}
\lref\Pasb{V.~Pasquier,
{\sl Incompressible representations of the Birman-Wenzl-Murakami algebra}, 
\pre{math.QA/0507364}.}
\lref\Jo{A. Joseph, {\sl On the variety of a highest weight module}, {\it J. Algebra} 88 (1) (1984), 238--278.}
\lref\Ro{W. Rossmann,
{\sl Equivariant multiplicities on complex varieties.
  Orbites unipotentes et repr\'esentations, III},
  {\it Ast\'erisque} No. 173--174, (1989), 11, 313--330.}
\lref\Ho{R.~Hotta, {\sl On Joseph's construction of Weyl group representations}, Tohoku Math. J. Vol. 36
(1984), 49--74.}
\lref\KZ{V.~Knizhnik and A. Zamolodchikov, {\sl Current algebra and Wess--Zumino model in two dimensions},
{\it Nucl. Phys.} B247 (1984), 83--103.}
\lref\FR{I.B.~Frenkel and N.~Reshetikhin, {\sl Quantum affine Algebras and Holonomic Difference Equations},
{\it Commun. Math. Phys.} 146 (1992), 1--60.}
\lref\JM{M.~Jimbo and T.~Miwa, {\it Algebraic analysis of Solvable Lattice Models}, 
CBMS Regional Conference Series in Mathematics vol. 85, American Mathematical Society, Providence, 1995.}
\lref\DFKZJ{P.~Di Francesco, A. Knutson and P. Zinn-Justin, work in progress.}
\lref\AS{F.~Alcaraz and Y.~Stroganov, {\sl The Wave Functions for the Free-Fermion Part of the Spectrum
of the $SU_q(N)$ Quantum Spin Models}, \pre{cond-mat/0212475}.}
\lref\DFZJc{P.~Di Francesco and P.~Zinn-Justin, 
{\sl Quantum Knizhnik--Zamolodchikov equation, generalized Razumov--Stroganov sum rules 
and extended Joseph polynomials}, 
{\it J. Phys. A} 38 (2005) L815--L822, \pre{math-ph/0508059}.}
\lref\DFZJd{P.~Di Francesco and P.~Zinn-Justin, 
{\sl From Orbital Varieties to Alternating Sign Matrices}, extended abstract for FPSAC'06,
\pre{math-ph/0512047}.}
\lref\DF{P.~Di Francesco,
{\sl Meander Determinants},
\pre{hep-th/9612026}.}
\lref\DFGG{P.~Di Francesco, O.~Golinelli and E.~Guitter,
{\sl Meanders and the Temperley--Lieb algebra},
{\it Commun. Math. Phys.} 186 (1997), 1--59,
\pre{hep-th/9602025}.}
\lref\KS{K.H. Ko and L.~Smolinksky,
{\sl A combinatorial matrix in 3-manifold theory},
{\it Pacific J. Math.} 149 (1991), 319--336.}
\lref\DFb{P.~Di Francesco,
{\it Inhomogeneous loop models with open boundaries},
{\it J. Phys.} A 38 (27) (2005), 6091--6120,
\pre{math-ph/0504032};
{\sl Boundary $q$KZ equation and generalized Razumov--Stroganov sum rules for open IRF models},
{\it J. Stat. Mech.} (2005) P11003,
\pre{math-ph/0509011}.}
\lref\Ma{P.~Martin,
{\sl Potts model and related problems in statistical mechanics},
World Scientific (1991).}
\lref\KS{W.M.~Koo and H.~Saleur,
{\sl Fused Potts models},
\pre{hep-th/930311}.}
\Title{}
{\centerline{Combinatorial point for higher spin loop models}
}
\bigskip\bigskip
\centerline{P. Zinn-Justin \footnote{${}^\star$}
{Laboratoire de Physique Th\'eorique et Mod\`eles Statistiques, UMR 8626 du CNRS,
Universit\'e Paris-Sud, B\^atiment 100,  F-91405 Orsay Cedex, France}}
\vskip1cm
\noindent Integrable loop models associated with higher representations (spin $\ell/2$) 
of $U_q(\goth{sl}(2))$ are investigated at the point $q=-\e{\pm i\pi/(\ell+2)}$. The ground
state eigenvalue and eigenvectors are described. Introducing inhomogeneities into the models allows to
derive a sum rule for the ground state entries.

\bigskip

\Date{03/2006}
%
%
%
\newsec{Introduction}
The present work is part of an ongoing project to understand the combinatorial properties of
integrable models at special points where a (generalized) stochasticity property is satisfied.
The project was started in \DFZJ, based on the observations and conjectures found in \refs{\BdGN,\RS}.
The original model under consideration was the XXZ spin chain with (twisted) periodic boundary conditions
at the special point $\Delta=-1/2$, or equivalently a statistical model 
of non-crossing loops with weight $1$ per loop (somewhat improperly called ``O(1)'' model, since it
is really based on $U_q(\widehat{\goth{sl}(2)})$ with $q=-\e{\pm i\pi/3}$), 
which can be reformulated as a Markov process on
configurations of arches. Among the 
various conjectured properties of the {\it ground state eigenvector},
a ``sum rule'' formulated in \BdGN, namely
that the sum of components of the properly normalized ground state eigenvector is equal
to the number of alternating sign matrices, was proved in \DFZJ.

Since then, a number of generalizations have been considered:
(i) models based on a different algebra, either the ortho/symplectic series which corresponds
to models of crossing loops \refs{\DFZJb,\KZJ}, or higher rank $A_n$ \DFZJc,
which can be described as paths in Weyl chambers. Note that in the latter case
the stochasticity property must be slightly modified: it becomes the existence of a (known) fixed
left eigenvector of the transfer matrix. This idea will reappear in the present work. 
(ii) models with other boundary conditions \refs{\DFb,\DFZJd}, which will not be discussed here.

There is yet another direction of generalization: the use of higher representations. Indeed
all models considered so far were based on fundamental representations (spin $1/2$
for $A_1$). We thus study here integrable models 
based on $A_1$, but representations of spin $\ell/2$. There is a reasonable way to formulate these
in terms of loops, using the fusion procedure (see Sect.~2).
One interesting feature is that the resulting models are closer
in their formulation to the original $O(1)$ loop model, 
and we can hope a richer combinatorial structure in the spirit
of the full ``Razumov--Stroganov conjecture'' \RS.

The present work remains indeed very close to that of \DFZJ. It is concerned with the study
of the ground state eigenvector and of the properties of its entries in an appropriate basis.
In fact, many arguments are direct
generalizations of those of \DFZJ\ -- though proofs are sometimes clarified and simplified.
There are however some new ideas. In particular, as already mentioned a key technical
feature
is the existence of a common left eigenvector for the whole family of operators from which one builds
the transfer matrix or the Hamiltonian. Here we give an ``explanation'' of this phenomenon: it is related
to the degeneration of a natural ``scalar product'' on the space of states. Indeed asking for this
scalar product to have rank 1 fixes the special value of the parameter $q$ to be $q=-\e{\pm i\pi/(\ell+2)}$,
which generalizes the value $\Delta={q+1/q\over2}=-1/2$ for spin 1/2. This will be explained in 
Sect.~3.1 and 3.2. Sect.~3.3 deals with the ground state eigenvector for
the inhomogeneous integrable transfer matrix, the polynomial character of
its components in terms
of the spectral parameters and other properties, while Sect.~3.4 describes the
computation of the sum rule, both following the general setup of \DFZJ.
In the latter, we shall be forced to rely on a conjecture concerning the degree of the polynomial
eigenvector: although in the special case $\ell=1$ this conjecture was proved in \DFZJ, 
the general proof is beyond the scope of the present paper.

\def\l{\kern.1em\ell\kern.1em}

\newsec{Definition of the model}
In this section we define the space of states and the Hamiltonian, or Transfer Matrix, acting on it.
In order to do that it is convenient to introduce a larger space, corresponding to the case $\ell=1$,
and then use fusion. This has the advantages that it gives us a natural ``combinatorial basis'' to work
with; however the situation, as we shall see, remains more subtle than in the case $\ell=1$, because
a projection operation is needed; in many cases, this means that results that are ``obvious graphically''
must be additionally shown to be compatible with the projection.

\subsec{Link Patterns and Temperley--Lieb algebra}
Let $n$ be a positive integer, and
$\LP$ be the set of {\it link patterns}\/ of size $n$,
which are defined as non-crossing (planar) pairings of $2n$ points. 
We want to imagine link patterns as living inside a disk,
with the $2n$ endpoints on the boundary; but it is sometimes more practical to unfold them to the traditional
depiction on a half-plane, see Fig.~\lppic. The number of such link patterns
is known to be the Catalan number $c_n=(2n)!/(n!(n+1)!)$. 
\fig\lppic{A link pattern.}%
{$\figbox{2.5cm}{lp-ex.eps}\qquad\to\qquad\figbox{5cm}{lp-ex2.eps}$}

We view $\LP$ as a subset of the involutions of $\{1,\ldots,2n\}$ without fixed points, by setting $\alpha(i)=j$
if $i$ and $j$ are paired by $\alpha\in\LP$.

Let $\H={\Bbb C}[\LP]$.
For $i=1,\ldots,2n-1$, we define $e_i$ to be the operator on $\H$ by defining its action on the canonical basis $\ket{\alpha}$,
$\alpha\in\LP$:
$$e_i \ket{\alpha}=\cases{\tau\ket{\alpha}&if $\alpha(i)=i+1$\cr \ket{c^{-1}\circ\alpha\circ c}&otherwise, $c$ cycle
$(i,i+1,\alpha(i+1),\alpha(i))$}
$$
where $\tau$ is a complex parameter, which for convenience we rewrite as 
$\tau=-q-q^{-1}$, $q\in{\Bbb C}^\times$.
We shall provide an alternative graphical rule below.

The $e_i$, $i=1,\ldots,2n-1$, form a representation of the usual Temperley--Lieb algebra $TL_{2n}(\tau)$.
By definition $TL_L(\tau)$ is the algebra with generators $e_i$, $i=1,\ldots,L-1$ and relations
\eqn\TLrel{
e_i^2=\tau e_i\qquad e_i e_{i\pm 1}e_i=e_i \qquad e_i e_j = e_j e_i\quad j\ne i-1,i+1\ .
}

It is well-known that the Temperley--Lieb algebra $TL_L(\tau)$ can be viewed itself as the space of linear combinations of 
non-crossing pairings of points on strips of size $L$, 
see the example below, multiplication being juxtaposition of strips, with the additional prescription
that each time a closed loop is formed, one can erase it at the price of multiplying by $\tau$.
In particular, the generators $e_i$ correspond to the strip
with two little arches connecting sites $i$ and $i+1$ on the top and bottom rows.
We conclude that the dimension of $TL_L(\tau)$ is $c_L$, so that for $L=2n$ it
is $c_{2n}=(4n)!/((2n)!(2n+1)!)$. The action on 
link patterns is once again juxtaposition of the strip and of the link pattern (in the unfolded
depiction), with the weight $\tau^{\#{\rm loops}}$ for erased loops.
Since $c_n^2<c_{2n}$, this representation is not faithful; however, when there is no possible confusion,
we shall by abuse of language identify Temperley--Lieb algebra elements and the corresponding operators on $\H$.

\example
$$TL_4=\Bigg\{\vtop{
\hbox{$
1=\figbox{2.4cm}{tl4-1.eps},\ e_1=\figbox{2.4cm}{tl4-12.eps},\ e_2=\figbox{2.4cm}{tl4-5.eps},
$}
\hbox{$
e_3=\figbox{2.4cm}{tl4-2.eps},\ e_1 e_2=\figbox{2.4cm}{tl4-11.eps},\ e_2 e_1=\figbox{2.4cm}{tl4-7.eps},
$}
\hbox{$
e_2 e_3=\figbox{2.4cm}{tl4-4.eps},\ e_3 e_2=\figbox{2.4cm}{tl4-3.eps},\ e_1 e_3=\figbox{2.4cm}{tl4-14.eps},
$}
\hbox{$
e_1 e_2 e_3=\figbox{2.4cm}{tl4-10.eps},\ e_3 e_2 e_1=\figbox{2.4cm}{tl4-6.eps},\ e_2 e_1 e_3=\figbox{2.4cm}{tl4-9.eps},
$}
\hbox{$
e_3 e_1 e_2=\figbox{2.4cm}{tl4-13.eps},\ e_2 e_1 e_3 e_2=\figbox{2.4cm}{tl4-8.eps}
\Bigg\}
$.}
}$$

In what follows, we shall sometimes need an extra operator $e_{2n}$, defined just like the other $e_i$, but reconnecting
the points $2n$ and $1$. The $e_i$, $i=1,\ldots,2n$ satisfy the same
types of relations \TLrel\ as before, but assuming periodic indices: $2n+1\equiv 1$.
These are defining relations of the ``periodic'' Temperley--Lieb
algebra $\widehat{TL}_{2n}(\tau)$. Clearly, its elements can be represented as
certain non-crossing pairings on an annular strip (acting in the obvious way on link patterns
in the circular depiction), but in practice are more complex to handle.
Fortunately in most circumstances we shall need to use only some subset 
of consecutive generators -- $e_i, e_{i+1},\ldots, e_{i+L-1}$, or $e_i,\ldots,e_{2n},e_1,\ldots, e_{i-2n+L-1}$ --
forming a representation of the usual (non-periodic) $TL_L(\tau)$.

\fig\bfpic{Gluing two link patterns together. Here two loops are formed.}{\epsfxsize=5cm\epsfbox{bf.eps}}
\subsec{Bilinear form}
There is an important pairing $\braket{\cdot}{\cdot}$ of link patterns which extends into a symmetric bilinear form on $\H$.
It consists of taking a mirror image of one link pattern, gluing it to the other and assigning it the usual weight
$\tau^{\#{\rm loops}}$, see Fig.~\bfpic. 
There is also an anti-automorphism $\ast$ of the periodic Temperley--Lieb algebra defined by 
$e_i{}_\ast=e_i$ (noting that the defining relations of $\widehat{TL}_{2n}(\tau)$ 
are invariant with respect to writing words in $e_i$ in the reverse order);
graphically, it associates to an operator its mirror image, and therefore we have the identity
\eqn\mir{
\bra{\alpha}x_\ast \ket{\beta}=\bra{\beta}x \ket{\alpha}\qquad x\in \widehat{TL}_{2n}(\tau)\ .
}
Define $g_{\alpha\beta}=\braket{\alpha}{\beta}$; the determinant of the matrix $g$ was computed in \DFGG,
In particular, it is non-zero when $q$ (that enters into the loop weight $\tau=-q-q^{-1}$) is generic,
i.e.\ not a root of unity (see also \KS).
However, in what follows we shall be particularly interested in 
the situation $q^{2(\ell+2)}=1$, in which $g$ is singular
for $n$ large enough, and
the mapping $\ket{\alpha}\mapsto \braket{\alpha}{\cdot}$ is {\it not}\/ an isomorphism from $\H$ to $\H^\star$, 
which requires some care in
handling bra-ket expressions. 

In particular, a remark is in order: in the ``strip'' description of the Temperley--Lieb algebra $TL_{2n}(\beta)$, 
it is clear
that any operator $\ket{\alpha}\braket{\beta}{\cdot}$ belongs to the Temperley--Lieb algebra (they are those pairings
of $2\times 2n$ points with no ``up--down'' pairings); therefore, for $q$ generic
the mapping from $TL_{2n}(\tau)$ to the space of operators $L(\H)$ is surjective. It is however in general
not surjective any more for $q$ root of unity; this is consistent with the fact that
there is no notion of adjoint operator with respect to the bilinear form for an arbitrary operator
on $\H$ (which $\ast$ provides for Temperley--Lieb elements), a point that will become crucial in Sec.~3.1.

\subsec{Projection}
Fix now a positive integer $\ell$, 
and assume that $n=\l m$. For each subset $S_i=\{\l(i-1)+1,\ldots,\l i\}$, $i=1,\ldots,2m$,
of $\ell$ consecutive points,
we define a local projector $p_i$; it is uniquely characterized by
\item{(i)} $p_i \ket{\alpha}=0$ if $\exists j,k\in S_i$ such that $\alpha(j)=k$.
\item{(ii)} $p_i$ is in the subalgebra generated by the $e_k$, $k=\l(i-1)+1,\ldots,\l i-1$.
\item{(iii)} $p_i^2=p_i$ (normalization).

The details of their construction and their main properties are listed in appendix A.
Here we give the key formula which is the recurrence definition: start with
$p^{(1)}=1$ and
\eqn\defproj{
p^{(k+1)}(e_j,\ldots,e_{j+k-1})
=p^{(k)}(e_j,\ldots,e_{j+k-2})(1-\mu_k(\tau) e_{j+k-1})
p^{(k)}(e_j,\ldots,e_{j+k-2})
}
where $\mu_k(\tau)=U_{k-1}(\tau)/U_k(\tau)$ and $U_k$ is the Chebyshev
polynomial of the second kind. Then
$p_i:=p^{(\ell)}(e_{\ell (i-1)+1},\ldots,e_{\ell i-1})$.

In particular we note that at the zeroes of the
Chebyshev polynomials $U_j(\tau)$, $1\le j\le \ell-1$,
that is if $q^{2j}= 1$ for some $1\le j\le \ell$, the $p_i$ are undefined;
we therefore exclude from now on these roots of unity.

The $p_i$ form a family of commuting orthogonal projectors; define $P=\prod_{i=1}^{2m} p_i$, and $\Hl=P(\H)$.
Furthermore, define
$$\LPl=\{\, \alpha\in\LP : \forall i, j\in S_i\ \alpha(j)\not\in S_i\,\}$$
that is, the set of link patterns with no arches within one of the subsets $S_i$.

\example
$${\cal L}_{2,6}=\Bigg\{\vtop{
\hbox{$
\figbox{2.2cm}{lp2-1.eps},\ 
\figbox{2.2cm}{lp2-2.eps},\ 
\figbox{2.2cm}{lp2-3.eps},\ 
\figbox{2.2cm}{lp2-4.eps},\ 
\figbox{2.2cm}{lp2-5.eps},
$}
\hbox{$
\figbox{2.2cm}{lp2-6.eps},\ 
\figbox{2.2cm}{lp2-7.eps},\ 
\figbox{2.2cm}{lp2-8.eps},\ 
\figbox{2.2cm}{lp2-9.eps},\ 
\figbox{2.2cm}{lp2-10.eps},
$}
\hbox{$
\figbox{2.2cm}{lp2-11.eps},\ 
\figbox{2.2cm}{lp2-12.eps},\ 
\figbox{2.2cm}{lp2-13.eps},\ 
\figbox{2.2cm}{lp2-14.eps},\ 
\figbox{2.2cm}{lp2-15.eps}
\Bigg\}$.}
}$$

It is crucial to observe that the
$\ket{\alpha}$, $\alpha\in\LPl$, {\it do not}\/ belong to $\Hl$. However, if we define $\ket{\alphal}:=P\ket{\alpha}$, $\alpha\in\LPl$, we can
state:

\prop\basis The $\ket{\alphal}$ form a basis of $\Hl$.

Proof. Clearly, $P\ket{\alpha}=0$ if $\alpha\not\in\LPl$.
Therefore $\dim\Hl\le\# \LPl$, and it suffices to show that the $\ket{\alphal}$ are independent.
But this is obvious in view of the fact that $\ket\alphal$ is of the form
$\ket{\alphal}=\ket{\alpha}+\sum_{\beta\not\in\LPl} c(\alpha,\beta)\ket{\beta}$ 
for all $\alpha\in\LPl$.
\qed

Note that this basis coincides with the dual canonical basis
of the invariant subspace of $2m$ copies of the $(\ell+1)$-dimensional 
representation of $U_q(\goth{sl}(2))$, see \Kho\ (thanks to K.~Shigechi for
pointing this out).

We can now introduce a set of local operators,
the $e_i^{(j)}$, $j=0,\ldots,\ell$, acting on the two subsets $S_i$ and $S_{i+1}$ for $i=1,\ldots,2m$
(with $S_{m+1}\equiv S_1$). They are defined by $e_i^{(j)}=P
\ e_{\ell i}\ e_{\ell i-1}e_{\ell i+1}\ \cdots\ e_{\ell i-j+1}\cdots e_{\ell i +j-1}\ \cdots
\ e_{\ell i-1}e_{\ell i+1}\ e_{\ell i}\ P$, but best understood graphically, see Fig.~\eijfig.

\fig\eijfig{Definition of $e_i^{(j)}$. The two subsets of vertices
are $S_i$ and $S_{i+1}$. The circled $p$'s are the 
local projection operators $p_i$ and $p_{i+1}$.}{$e^{(j)}=\figbox{4.5cm}{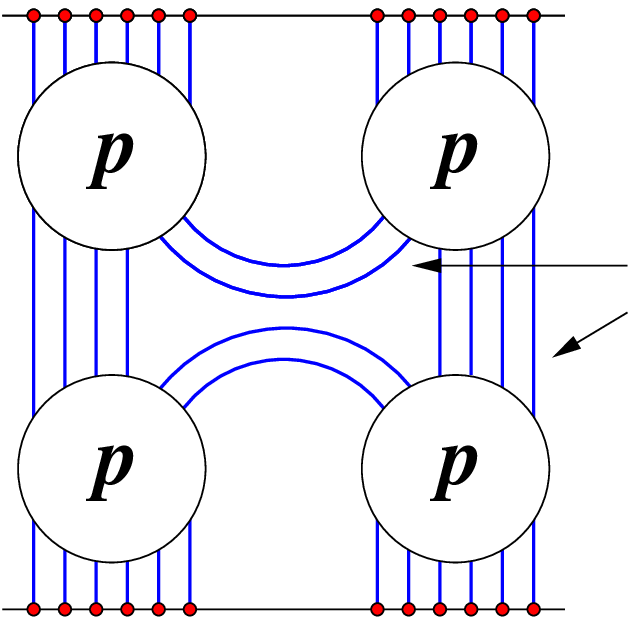}$
\vbox{\hbox{$j$ lines}\hbox{$\ell-j$ lines}}
}


The $e_i^{(j)}$ satisfy a certain algebra which we do not need to describe entirely. 
However we need the following results:

\lemma\eij The image of $e_i^{(j)}$ is included in the span of the
$\ket\alphal$ such that there are (at least)
$j$ arches between $S_i$ and $S_{i+1}$,
i.e.\ $\alpha(\ell i)=\ell i+1$, \dots,
$\alpha(\ell i-j+1)=\ell i+j$.

Proof. This
is obvious graphically, since $e_i^{(j)}$ reconnects precisely these pairs of points mentioned in the lemma, then projects. \qed

\lemma\eijb The equality of Fig.~\eijbfig\ holds. Consequently,
(a) Consider a link pattern $\alpha$ such that $S_i$ and $S_{i+1}$ are fully connected, i.e.\ 
$\alpha(\ell i)=\ell i+1$, \dots,
$\alpha(\ell (i-1)+1)=\ell (i+1)$. Then
\eqn\eijba{
e_i^{(j)} \ket\alphal= {U_\ell(\tau)\over U_{\ell-j}(\tau)} \ket\alphal\ .
}
(b) Equivalently,
\eqn\eijbb{e_i^{(j)}e_i^{(\ell)}=e_i^{(\ell)}e_i^{(j)}=
{U_\ell(\tau)\over U_{\ell-j}(\tau)}e_i^{(\ell)}\ .}

\fig\eijbfig{Graphical equality of Lemma~\eijb. $p$ refers to 
$p^{(\ell)}$ in the l.h.s.\ and to $p^{(\ell-j)}$ in the r.h.s.}{%
\vbox{\hbox to1.6cm{\hfil\rm$j$ lines}\hbox to1.6cm{\hfil\rm$\ell-j$ lines}}\ 
$\figbox{4cm}{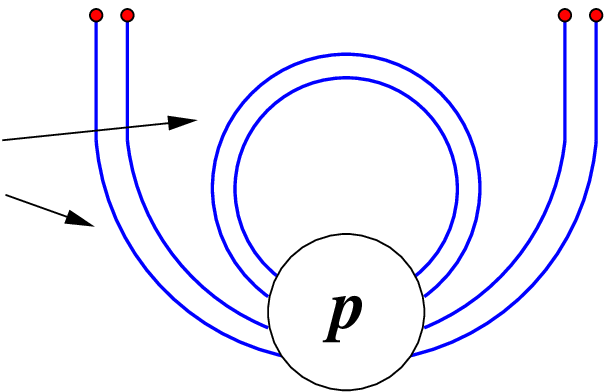}={\displaystyle U_\ell(\tau)\over\displaystyle U_{\ell-j}(\tau)}
\figbox{4cm}{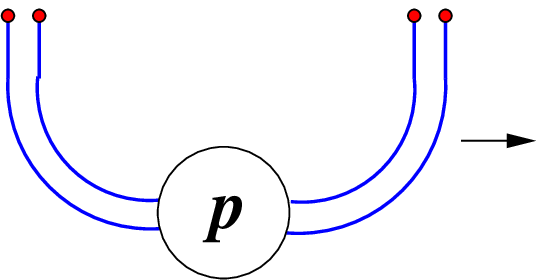}$ $\ell-j$ lines}

Proof. The proof is by induction on $\ell$. Consider the l.h.s.\ of Fig.~\eijbfig\ and replace
$p^{(\ell)}$ with its definition by recurrence from Appendix A, choosing
to apply $p^{(\ell-1)}$ to the $\ell-j$ open lines and to $j-1$ closed lines,
excluding the innermost closed line: we obtain two terms
which are both precisely of the same form as the l.h.s., 
but with $\ell-1$ lines among which $j-1$ close,
and coming with coefficients $\tau$ (one closed loop) and $-\mu_{\ell-1}(\tau)$.
Applying the induction hypothesis we find the coefficient of proportionality to be
$(\tau-U_{\ell-2}(\tau)/U_{\ell-1}(\tau))U_{\ell-1}(\tau)/U_{\ell-j}(\tau)
=(\tau U_{\ell-1}(\tau)-U_{\ell-2}(\tau))/U_{\ell-j}(\tau)=U_{\ell}(\tau)/U_{\ell-j}(\tau)$. 
The proof of Eqs.~\eijba\ and \eijbb\ is a simple application of this formula, noting
that when there are series of projections one can coalesce them into a single projection.
\qed

This second lemma is particularly important; 
we provide on Fig.~\eijbfigg\ two more graphical corollaries of it.
\fig\eijbfigg{Equalities obtained from Lemma~\eijb. On the two figures $p= p^{(\ell)}$.}{(a)
\vbox{\hbox to1.6cm{\hfil $j$ lines}\hbox to1.6cm{\hfil$\ell-j$ lines}}\ 
$\figbox{4cm}{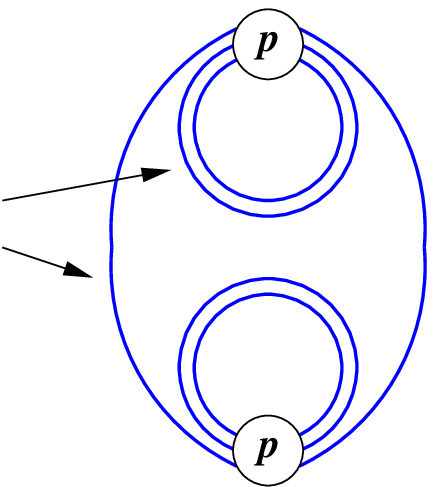}
={\displaystyle U_\ell(\tau)^2\over\displaystyle U_{\ell-j}(\tau)}$\qquad
(b)
$\figbox{2cm}{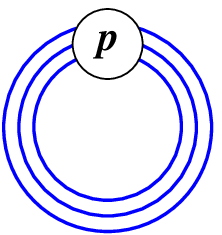}
=U_\ell(\tau)$
}

\subsec{Fusion}
Let us briefly describe the fusion mechanism. Since this is standard material,
we shall not prove the following facts.

Start with the $\ell=1$ $R$-matrix
\eqn\defr{
r_i(z,w)={q z -q^{-1} w\over q w-q^{-1} z} +{z-w\over q w-q^{-1} z}e_i\ .
}

We now fuse $\ell^2$ $R$-matrices\foot{Note that to define a transfer matrix, one could fuse only
$\ell$ $R$-matrices, keeping a single line for the ``auxiliary space''. However we need
the doubly fused $R$-matrix to write the
form of the Yang--Baxter equation that we need, and to obtain the Hamiltonian.} into a single operator $R$ by 
\eqnn\defR
$$\eqalignno{
R_i(z,w)=&\left(\prod_{k=1}^\ell {q^{-k}z-q^k w\over q^k z-q^{-k}w}\right)
\ r_{\ell i}(q^{-\ell+1}z,q^{\ell-1} w)\cr
&r_{\ell i-1}(q^{-\ell+3}z,q^{\ell-1} w)r_{\ell i+1}(q^{-\ell+1}z,q^{\ell-3}w)\cr
&\cdots\cr
&r_{\ell(i-1)+1}(q^{\ell-1} z,q^{\ell-1} w)\cdots r_{\ell(i+1)-1}(q^{-\ell+1} z,q^{-\ell+1}w)\cr
&\cdots\cr
&r_{\ell i-1}(q^{\ell-1} z,q^{-\ell+3}w)r_{\ell i+1}(q^{\ell-3}z,q^{-\ell+1} w)\cr
&r_{\ell i}(q^{\ell-1} z,q^{-\ell+1}w)\ P\ .&\defR\cr}
$$
Due to the choice of arguments of the $R$-matrices, $R_i$ leaves $\Hl$ stable.

We shall need a more explicit form of $R_i$. 
This is possible, using the local operators $e_i^{(j)}$ introduced previously:
\eqn\defRb{
R_i(z,w)=\sum_{j=0}^\ell a_j \left(\prod_{k=0}^{\ell-j} {q^{k} z-q^{-k}w\over q^{k-\ell}z-q^{\ell-k}w}\right)
e_i^{(j)}
}
where $a_j=a_{\ell-j}=\prod_{k=1}^j {U_{\ell-k}(\tau)\over U_{k-1}(\tau)}$.

$R_i(z,w)$ has poles when $w/z=q^{-2},\ldots,q^{-2\ell}$ and is non-invertible
when $w/z=q^2,\ldots,q^{2\ell}$; for other values of $z/w$ it
satisfies the unitarity equation
\eqn\uni{
R_i(z,w)R_i(w,z)=1\ .
}

Next we define the fully inhomogeneous transfer matrix $T(z| z_1,\ldots,z_{2m})$.
This requires to extend slightly the space $\Hl$ into $\Hlaux$ where the additional ``auxiliary''
$\ell$ lines are drawn horizontally.
One also defines the ``partial trace'' $\tr_{aux}$ which
to an operator on $\Hlaux$ associates an operator on $\Hl$ obtained by reconnecting together
the incoming and outgoing auxiliary lines,\foot{Making the auxiliary line horizontal
conceals the fact that this operation induces a rotation of the link pattern, since the
auxiliary line has changed its position from left to right relative to the rest of the lines.}
including a weight of $\tau=-q-q^{-1}$ by closed loop.
Then the transfer matrix corresponds to the auxiliary line crossing all other lines
then reconnecting itself (see Fig.~\tm)
\eqn\defT{
T(t|z_1,\ldots,z_{2m})=\tr_{aux} R_{2m}(z_{2m},t)\cdots R_2(z_2,t)R_1(z_1,t) 
}
\fig\tm{The transfer matrix ($\ell=3$, $2m=4$). Each $\ell\times\ell$ grid is the fused $R$-matrix.}{%
\epsfxsize=4cm\epsfbox{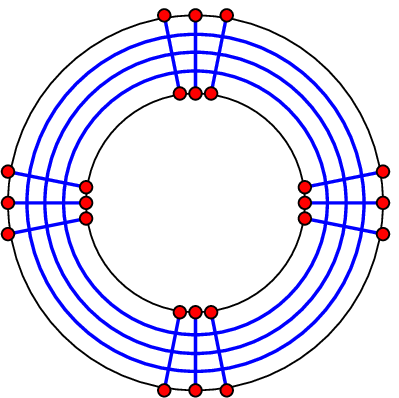}}

The transfer matrix satisfies two forms of the Yang--Baxter equation. The first one is the well-known
``RTT'' form, which implies the commutation relation $[T(t),T(t')]=0$,
where all $z_i$ are fixed. The second one simply reads:
\eqn\YBE{
T(t|z_1,\ldots,z_{i},z_{i+1},\ldots,z_{2m})
R_i(z_i,z_{i+1})
=R_i(z_i,z_{i+1})
T(t|z_1,\ldots,z_{i+1},z_{i},\ldots,z_{2m})
}
for $i=1,\ldots,2m$ (with $z_{2m+1}\equiv z_1$).

We also need the ``scattering matrices'' $T'_i:=T(z_i|z_1,\ldots,z_{2m})$, $1\le i\le 2m$.
Using the fact that $R_i(z,z)=1$, one finds
\eqn\scatt{
T'_i=R_i(z_i,z_{i+1})\ldots R_{2m-1}(z_i,z_{2m})R_{2m}(z_i,z_1)R_1(z_i,z_2)\ldots R_{i-1}(z_i,z_{i-1})\rho}
where $\rho$ is the rotation of link patterns:
$(\rho\alpha)(i)=\alpha(i-\ell)+\ell$ (modulo $2n$),
which sends $S_i$ to $S_{i+1}$.

One can also define the Hamiltonian. Consider the homogeneous situation $z_i=1$.
Then it is natural to expand $T$ around $t=1$ to obtain commuting operators
that are expressed as sums of local operators (the $e_i^{(j)}$). Explicitly, expanding at
first order, we find that $T(t)$ commutes with
\eqn\defH{
H:=(q-q^{-1})\,T(1)^{-1}{\der\over\der t}T(t)|_{t=1} +cst
=\sum_{i=1}^{2m}\sum_{j=1}^\ell {1\over U_{j-1}(\tau)} e^{(j)}_i
}
(the constant term in the expansion has been cancelled for convenience).

\subsec{Cell depiction}
Finally,
there is yet another graphical depiction of link patterns in $\LPl$: since vertices in the same subset $S_i$ are never
connected to each other, one can simply coalesce them into a single vertex: the result is a division of the disk into 2-dimensional cells such that $\ell$ edges come out of each of the $2m$ vertices on the boundary.

\example 
at $\ell=2$ cells can be conveniently drawn using
the natural bicoloration of cells according to whether they touch the exterior circle
at vertices or edges (see also below the discussion of exterior vs interior cells):
$\figbox{2.2cm}{lp2-4.eps}=\figbox{2.2cm}{lp2b-4.eps}$. 
Note that if one straightens edges to produce polygons, one can obtain 2-gons (or worse, several 2-gons that sit
on top of each other); it is
therefore possible to work with polygonal cells on condition that such singular configurations be included.

For future use, we now define the following notion:
a link pattern $\alpha\in\LPl$ is said to be {\it $\ell$-admissible}\/ 
if all its cells have an even number of edges.
When there is no ambiguity we shall simply say ``admissible'', noting
that this an abuse of language since admissibility
is an $\ell$-dependent property: some edges disappear when vertices are merged.
Call $\LPl'$ the set of $\ell$-admissible link patterns.

\example
$${\cal L}'_{2,6}=\Bigg\{\vtop{
\hbox{$
\figbox{2.2cm}{lp2b-1.eps},\ 
\figbox{2.2cm}{lp2b-2.eps},\ 
\figbox{2.2cm}{lp2b-3.eps},\ 
\figbox{2.2cm}{lp2b-5.eps},\ 
\figbox{2.2cm}{lp2b-6.eps},
$}
\hbox{$
\figbox{2.2cm}{lp2b-7.eps},\ 
\figbox{2.2cm}{lp2b-8.eps},\ 
\figbox{2.2cm}{lp2b-9.eps},\ 
\figbox{2.2cm}{lp2b-11.eps},\ 
\figbox{2.2cm}{lp2b-13.eps},
$}
\hbox{$
\figbox{2.2cm}{lp2b-14.eps},\ 
\figbox{2.2cm}{lp2b-15.eps}
\Bigg\}$.}
}$$

We also need a simple fact about admissible link patterns.
Call $r(i)$ the remainder of the division of $i-1$ by $\ell$.

\lemma\parity If $\alpha$ is an $\ell$-admissible link pattern,
then $r(i)+r(\alpha(i))=\ell-1$ for $1\le i\le 2m$.

Proof. Induction on $\alpha(i)-i\ ({\rm mod}\ 2m)\, \in\{1,\ldots,2m-1\}$ .

$\star$ If $\alpha(i)=i+1$: $\alpha\in {\cal L}_{\ell,2m}$ forbids any arches inside
a given subset of $\ell$ vertices, therefore $r(i)=\ell-1$, $r(i+1)=0$.

$\star$ If $\alpha(i)-i>1$:  call $k=r(i+1)$. 
Consider the cell with edge $(i,\alpha(i))$ such that all its other vertices are between $i$ and
$\alpha(i)$
moving counterclockwise around the circle.
The idea is to use the induction hypothesis for all 
these other vertices.
Two cases have to be distinguished:
either (i) $k=0$, in which case the values
of $r$ at vertices (in the sense of the original depiction) of the cell follow a pattern:
$0$, $\ell-1$, $0$, etc and we obtain immediately $r(i)=\ell-1$, $r(\alpha(i))=0$; or (ii) $k>0$. 
In this case, the values of $r$ at vertices of the cell are of the form
$k$, $\ell-1-k$, $\ell-k$, $k-1$, $k$, etc, being careful that these vertices are coalesced into
pairs $\{k-1,k\}$ and $\{ \ell-1-k,\ell-k\}$ to form the actual vertices of the cell.
But since $\alpha$ is admissible, the cell has an even number of edges,
and when we reach $\alpha(i)$ we get the value $\ell-k$, so that
$r(i)=k-1$, $r(\alpha(i))=\ell-k$. \qed

In the course of the proof, we have found that one can
associate to each cell $c$ of an admissible link pattern a pair of integers $\{ k(c), \ell-k(c) \}$: 
conventionally we choose $k(c)$ to be the smaller of the two.
Graphically, $k(c)$ is
the ``distance'' from the cell to the boundary, defined as the
minimum number of edges one needs to cross to reach the exterior circle (excluding the circle itself). 
Following the subdivision in the proof,
We call {\it exterior}\/ (resp.\ {\it interior}) a cell $c$ such that $k(c)=0$ (resp.\ $k(c)>0$).
An exterior cell touches the circle at every other edge, whereas an interior cell touches it at vertices
only.
In practice exterior cells play no role in what follows,
as will become clear, and on the pictures they will be left uncolored.
Note that in the case $\ell=2$ this notion coincides with the natural bicoloration of cells.

Also note that the converse of lemma \parity\ is untrue. In particular if $\ell=2$
the property of lemma \parity\ is always satisfied by parity.

In appendix B, $\ell$-admissible link patterns are enumerated, and it is found that
\eqn\enumadm{
\#\LPl'={((\ell+1)m)!\over (\l m+1)!m!}\ .
}

\newsec{Combinatorial point}
We now investigate the special value
$q=-\e{\pm i\pi/(\ell+2)}$, that is
$\tau=-q-q^{-1}=2 \cos{\pi\over\ell+2}=1,\sqrt{2},{1+\sqrt{5}\over 2},\sqrt{3},\ldots$.

\goodbreak
\subsec{Degeneration of the bilinear form}
Define the matrix of the bilinear form in the subspace $\Hl$:
$$\tilde g_{\alpha\beta}:=\braket{\alphal}{\betal}=\bra{\alpha}P\ket{\beta}
\qquad \alpha,\beta\in\LPl\ .$$

\thm\rankg The rank of the matrix $\tilde g$ is one.

Proof. As many reasonings in this paper, the proof is best
understood pictorially. It makes use of Lemma~\eijb, with the additional assumption
that $q=-\e{\pm i\pi/(\ell+2)}$, which implies that $U_j(\tau)=U_{\ell-j}(\tau)$.
Fig.~\eijbfig(a) therefore implies the equality of Fig.~\rankfig(a), 
which itself can be rewritten as Fig.~\rankfig(b),
noting that any link pattern in ${\cal L}_{\ell,4}$ is of the form of Fig.~\rankfig(a) for some $j$
-- and for any other link pattern in ${\cal L}_{4\ell}$, both l.h.s.\ and r.h.s.\ are zero.
\fig\rankfig{Graphical proof of Thm.~\rankg. L stands for any linear combination of link
patterns.}{%
\vbox{
\hbox{(a) \vbox{\hbox to1.6cm{\hfil $j$ lines}\hbox to1.6cm{\hfil$\ell-j$ lines}}
$\figbox{3.5cm}{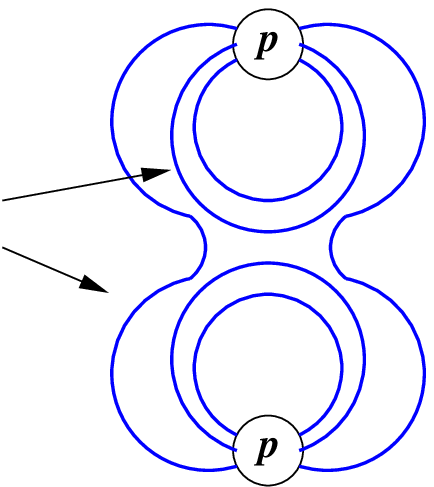}
=\ \figbox{4cm}{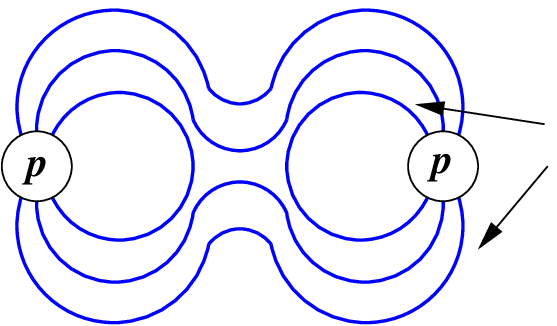}$
\vbox{\hbox{$\ell-j$ lines}\hbox{$j$ lines}}}
\hbox{(b) $\figbox{2cm}{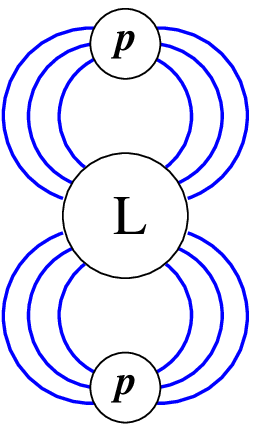}=\figbox{3.5cm}{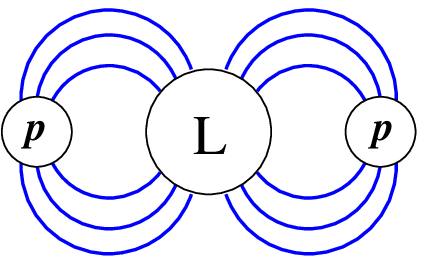}$}
\vskip8pt
\hbox{(c) $\figbox{4.5cm}{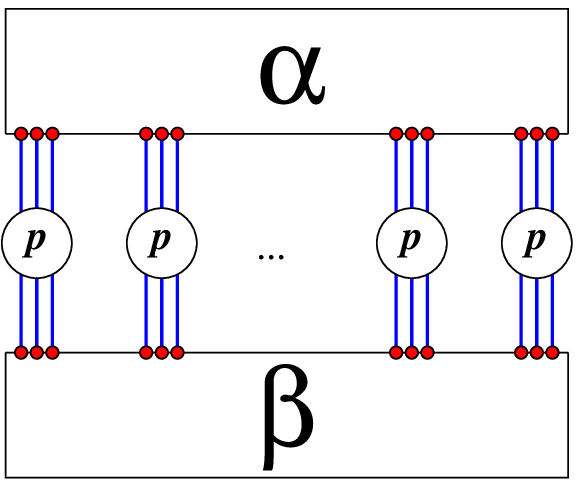}=\figbox{4.5cm}{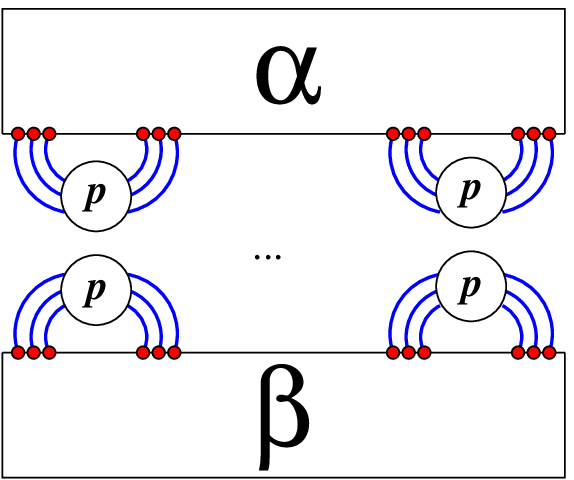}$}
}
}
Consider now the braket $\bra{\alpha}P\ket{\beta}$. Using repeatedly the identity of Fig.~\rankfig(b),
we obtain Fig.~\rankfig(c), that is
\eqn\rankone{
\braket{\alphal}{\betal}=\braket{\alphal}{0}\braket{0}{\betal}
}
where $0$ denotes the link pattern 
which fully connects $S_{2i-1}$ and $S_{2i}$, as in the r.h.s.\
of Fig.~\rankfig(c) (e.g.\ $\figbox{2cm}{lp3-1.eps}$). 
Thus, $\tilde g=v\otimes v$, $v$ the linear form $\braket{0}{\cdot}$ on $\Hl$
which is non-zero since $\tilde{g}_{00}=\bra{0}P\ket{0}=1$ (Fig.~\eijbfig(b)). \qed

Remark: there is another link pattern $0'$ related to $0$ by rotation, which connects
$S_{2i}$ and $S_{2i+1}$ ($2m+1\equiv 1$). The argument above works equally well with $\bra{0'}$.

We can in fact provide an explicit formula for $\tilde g$, of the form
$\tilde g_{\alpha\beta}= v_\alpha v_{\beta}$, $\alpha,\beta\in\LPl$:
\prop\explv
\eqn\defv{
v_\alpha = \braket{0}{\alphal}=\cases{0&if $\alpha$ is non-admissible\cr 
\prod_{{\rm cell}\ c} U_{k(c)}(\tau)^{-l(c)/2+1}
&if $\alpha$ is admissible}
}
where the product can be restricted to interior cells only,
$l(c)$ is the number of edges of cell $c$ (note that 2-gons
do not contribute), and $k(c)$ is the distance from the cell to the boundary as
defined in Sect.~2.3.

Proof. Induction on $m$. $m=1$ is trivial. For a given link pattern $\alpha$,
we shall pick a certain pair of subsets $S_i$, $S_{i+1}$ and reconnect them with a projection: this
is one step in the pairing with $\bra{0}$ (or $\bra{0'}$, depending on the parity of $i$), and we
can then use the induction hypothesis.

For any link pattern, it is easy to check that one of these two situations must arise
(graphically, that there exists a cell which has no ``nested'' cells):

(i) either there are two subsets $S_i$ and $S_{i+1}$ which are fully connected to each other.
These correspond to $\ell$ 2-gons which should not contribute to $v_\alpha$. Indeed, applying
Fig.~\eijbfigg(b), the loops, once closed with a projection, contribute $U_\ell(\tau)=1$ and can be
removed, leading to the step $m-1$.

(ii) or there is a subset $S_i$ such that $j$ lines connect it to $S_{i-1}$ and $\ell-j$ lines
connect it to $S_{i+1}$. We reconnect $S_i$ and $S_{i+1}$ and apply Lemma~\eijb\ (Fig.~\eijbfig).
In the process some 2-gons are erased, and the only other (interior) cell that is affected is
the one directly above, see Fig.~\recv, which we denote by $c$. One checks that $c$ loses 2 edges. 
If $c$ had 3 edges to begin with, it becomes a
cell with 1 edge i.e.\ there is a connection inside a subset and the resulting link pattern
does not belong to ${\cal L}'_{\ell,2(m-1)}$, so $v_\alpha=0$. If $c$ had a higher odd number of edges
the resulting link pattern is not admissible and by induction $v_\alpha$ is again zero. Finally, if the
number of edges of $c$ is even,
we note that $\min(j,\ell-j)=k(c)$ and the contribution $1/U_j(\tau)$ plus the
induction hypothesis reproduce Eq.~\defv\ (whether $\alpha$ is admissible or not). \qed
\fig\recv{Case (ii) of the proof of Prop.~\explv.}{%
$\figbox{4cm}{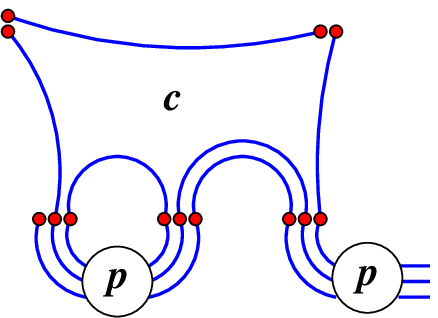}={\textstyle 1\over\textstyle U_{k(c)}(\tau)}\figbox{3.4cm}{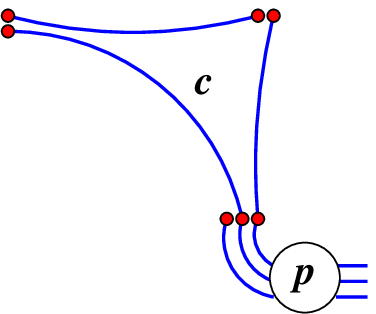}$}

\example consider the diagram $\alpha=\figbox{2.2cm}{lp3b-ex.eps}\in {\cal L}_{3,10}$. It is admissible,
and there are two 4-gons at distance 1, so $v_\alpha=U_1(\tau)^{-2}=((1+\sqrt{5})/2)^{-2}$.\hfil\break
$\alpha=\figbox{2.2cm}{lp4b-ex.eps}\in {\cal L}_{4,10}$ is also admissible,
there are 4-gons at distance 1 and 2, so $v_\alpha=U_1(\tau)^{-1}U_2(\tau)^{-1}=1/(2\sqrt{3})$. 

Remark. For $\ell=2,3$, since the only non-trivial $U_j(\tau)$ are equal to $\tau$,
one can simplify the formula for admissible link patterns to:
$v_\alpha=\tau^{-m+\#\hbox{connected components of cells}}$.

\subsec{Common left eigenvector}
Consider now any operator $x$ of the (periodic) Temperley--Lieb algebra, projected onto $\Hl$, 
that is $x=PxP$.
As explained in Sec.~2.2, it possesses a mirror symmetric $x_\ast$.
Let us write in components the identity \mir\ expressing this fact: if $x\ket{\alphal}=\sum_{\beta}x^{\beta}_{\ \alpha} \ket{\betal}$
and $x_\ast\ket{\alphal}=\sum_{\beta}x^\beta_{\ast\alpha} \ket{\betal}$, then
$\tilde g_{\alpha\gamma}x^\gamma_{\ast\beta}=\tilde g_{\beta\gamma}x^\gamma_{\ \alpha}$, where summation over
repeated indices is implied, or, choosing any $\beta$ such that $v_\beta\ne0$,
\eqn\miracle{
v_\gamma\, x^\gamma_{\ \alpha}=
{v_\gamma x^\gamma_{\ast\beta} \over v_\beta}\, v_\alpha\ .
}
In other words, $v$ is a left eigenvector of $x$ (and of $x_\ast$ by exchanging their roles).
What we have found is that the right-representation of $P\, \widehat{TL}_{2n}(\tau)P$ on $\Hl^\star$ possesses a one-dimensional
stable subspace; and therefore also that the left-representation on $\Hl$ is decomposable (but not reducible, as it turns out)
with a stable subspace of codimension one (the kernel of $v$).
Note that an advantage of defining the left eigenvector $v$ from the bilinear form is that
it provides a convenient natural normalization of $v$.

\eqnn\vev
\lemma\explev Eigenvalues of various operators for the left eigenvector $v$:
$$\eqalignno{
v\, e_i^{(j)}&={1\over U_j(\tau)} v\qquad j=0,\ldots,\ell\cr
v\,  R_i(z,w)=v\cr
v\, T(t|z_1,\ldots,z_{2m})&=v\cr
v\, H&=2m\tau v&\vev\cr
}$$

Proof. Since we already know that $v$ is a left eigenvector
of $e_i^{(j)}$, we only need to compute $v\, e_i^{(j)} \ket{\alphal}$ where $\alpha$ 
is a given admissible link
pattern; we choose it as in the hypotheses of Lemma~\eijb\ 
(for example, either $\ket{0}$ or $\ket{0'}$ works).
We conclude directly that $U_{\ell}(\tau)/U_{\ell-j}(\tau)$ is the eigenvalue
for $v$, which is the announced result using $U_j(\tau)=U_{\ell-j}(\tau)$ at $q=-\e{\pm i\pi/(\ell+2)}$. 
The other formulae follow by direct computation. \qed

\lemma\simpleev The eigenvalue $1$ of $T(t|z_1,\ldots,z_{2m})$ is simple for generic values of the parameters.

Note that the set of degeneracies of the eigenvalue $1$ is a closed subvariety of the space of parameters.
Thus, finding one point where the eigenvalue is simple is enough to show the lemma.
There are a variety of ways to find such a point, none of which being particularly simple.
One can for example consider the limit $z_1\ll z_2\ll\cdots\ll z_{2m}$, in which
all eigenvalues can be computed explicitly.
The calculations are too cumbersome and will not be reproduced here.

\subsec{Polynomial eigenvector}
We have found in the previous section that the transfer matrix
$T(t|z_1,\ldots,z_{2m})$ possesses the eigenvalue $1$, with left eigenvector $v$; what about
the corresponding right eigenvector? The latter, which we denote by 
$\ket\Psi=\sum_\alpha \Psi_\alpha \ket{\alphal}$, depends on the parameters
$z_1,\ldots,z_{2m}$ (but not on $t$). Being the solution of a degenerate linear
system of equations whose coefficients are rational fractions, it can be normalized in such a way
that its components $\Psi_\alpha$
are {\it coprime polynomials}\/ in the variables $z_1,\ldots,z_{2m}$.
Furthermore, all equations being homogeneous, the $\Psi_\alpha$ are homogeneous polynomials
of the same degree $\deg \ket\Psi$. We now formulate a key result:

\prop\exch $\ket{\Psi(z_1,\ldots,z_i,z_{i+1},\ldots,z_{2m})}=
R_i(z_i,z_{i+1}) \ket{\Psi(z_1,\ldots,z_{i+1},z_i,\ldots,z_{2m})}$.

Proof. 
Eq.~\YBE\ shows that $R_i(z_i,z_{i+1})\ket{\Psi(z_{i+1},z_i)}$ is an eigenvector of $T(z_1,\ldots,z_{2m})$
with the eigenvalue $1$ (Lemma \explev). Since this eigenvalue is simple (Lemma \simpleev),
the l.h.s.\  and r.h.s.\ of Prop.~\exch\ must be proportional: 
$R_i(z_i,z_{i+1}) \ket{\Psi(z_{i+1},z_i)}
=F(z_1,\ldots,z_{2m})\ket{\Psi(z_i,z_{i+1})}$, $F$ rational fraction.
More precisely,
\eqn\protoexch{
\prod_{k=1}^\ell (q^{-k} z_i-q^{k}z_{i+1})
R_i(z_i,z_{i+1}) \ket{\Psi(z_{i+1},z_i)}
=P(z_1,\ldots,z_{2m})\ket{\Psi(z_i,z_{i+1})}
}
where $P(z_1,\ldots,z_{2m})=F(z_1,\ldots,z_{2m})\prod_{k=1}^\ell (q^{-k} z_i-q^{k}z_{i+1})$
is a polynomial since the $\Psi_\alpha$ are coprime and the l.h.s.\ is already polynomial. 
Iterating this equation leads to
\eqn\condP{
P(z_i,z_{i+1})P(z_{i+1},z_i)=\prod_{k=1}^\ell (q^{-k} z_i-q^{k}z_{i+1})(q^{-k} z_{i+1}-q^{k}z_{i})
}
i.e.\ $P$ and therefore $F$ are functions of only two variables and
$F(z,w)=\prod_{k\in K} {q^{-k} w-q^k z\over q^{-k} z-q^k w}$
where $K$ is some subset of $\{1,\ldots,\ell\}$. To fix $F$,
consider $T'_i \ket{\Psi}$ (with $T'_i$ given by Eq.~\scatt): 
on the one hand, since $T'_i$ is simply the transfer matrix at a special
choice of parameter $t$, we know that it has
eigenvector $\Psi$ with eigenvalue $1$
(Lemma \explev): $T'_i\ket{\Psi}=\ket{\Psi}$; 
on the other hand, applying Eq.~\protoexch\ repeatedly,
we find that $T'_i\ket{\Psi}=\prod_{j\ne i} F(z_i,z_j) \ket{\Psi}$; 
we easily conclude from this that
$F=1$. 
\qed

\prop\littlecancel Suppose $z_{i+1}=q^{2k} z_i$, $1\le k\le \ell$.
Then $\Psi_\alpha(z_1,\ldots,z_{2m})=0$ unless $\alpha$ is such that
there are (at least) $\ell-k+1$ arches between $S_i$ and $S_{i+1}$.

Proof. Apply proposition \exch\ with $R_i(z_i,q^{2k}z_i)$ replaced with its expression $\defRb$.
As soon as $\ell-j\ge k$, the product is zero, so that $R_i$ is a linear combination
of $e_i^{(\ell-k+1)}$, \dots, $e_i^{(\ell)}$. The proposition is then a direct application of Lemma~\eij.
\qed

Equivalently, since the components $\Psi_\alpha$ are polynomials, 
$z_{i+1}-q^{2k} z_i | \Psi_\alpha$. We now state a broad generalization of Prop.~\littlecancel:

\thm\bigcancel Suppose $z_j=q^{2k} z_i$, $1\le k\le \ell$, $j\ne i$.
Then $\Psi_\alpha(z_1,\ldots,z_{2m})=0$ unless $\#\{ p,q\in S_{i,j} : p<\alpha(p)=q \}\ge \ell-k+1$.

Here $S_{i,j}$ denotes the set of vertices between $i$ and $j$ in a cyclic way, that is
$S_{i,j}=\{ \ell(i-1)+1,\ldots, \ell j\}$ if $i<j$,
$\{ \ell(i-1)+1,\ldots,2m,1,\ldots,\ell j\}$ if $j<i$.

Note that this theorem is a 
generalization of (part of) theorem 1 of \DFZJ, and the proof is completely analogous.
We present here a briefer version of it. 
Prop.~\exch\ shows that $\ket{\Psi(\ldots,z_i,z_{i+1},\ldots,z_j,\ldots)}$ and 
$\ket{\Psi(\ldots,z_i,z_j,z_{i+1},\ldots)}$ are related by a product of $R$-matrices
from $i+1$ to $j-1$
(these $R$-matrices have poles, but are well-defined for generic values of the other $z$'s).
According to Prop.~\littlecancel, the only non-zero components
of $\ket{\Psi(\ldots,z_i,z_j,z_{i+1},\ldots)}$ possess $\ell-k+1$ arches between subsets $S_i$ and $S_{i+1}$.
Now observe that the action of any Temperley--Lieb generator {\it cannot decrease}\/ the number of arches
within any range containing the 2 sites on which it is acting. Therefore multiplication by
a product of $R$-matrices (which are themselves linear combinations 
of products of Temperley--Lieb generators
acting somewhere between $S_i$ and $S_j$)
does not decrease the number of arches between $S_i$ and $S_j$. \qed

One can go further and look for cancellation conditions for the whole of $\ket\Psi$:

\prop\littlecancelglob Assume that $z_{i+1}=q^{2k} z_i$ and $z_{i+2}=q^{2k'} z_{i+1}$
so that $z_i,z_{i+1},z_{i+2}$ are in ``cyclic order'',
that is $1\le k,k'$ and $k+k'\le \ell+1$. Then $\ket{\Psi(z_1,\ldots,z_{2m})}=0$.

Proof. apply twice Prop.~\littlecancel. For a component $\Psi_\alpha$ to be non-zero,
$\alpha$ should have $\ell-k+1$ arches between $S_i$ and $S_{i+1}$,
and $\ell-k'+1$ arches between $S_{i+1}$ and $S_{i+2}$; thus the number of lines emerging from
$S_{i+1}$ should be $2\ell-(k+k')+2\ge \ell+1$, which is impossible. \qed

Once again we can generalize this result to
\thm\bigcancelglob Assume that $z_{i'}=q^{2k} z_i$ and $z_{i''}=q^{2k'} z_{i'}$
so that $z_i,z_{i'},z_{i''}$ are in cyclic order,
and $i,i',i''$ are also in cyclic order ($i<i'<i''$ or $i'<i''<i$ or $i''<i<i'$). 
Then $\ket{\Psi(z_1,\ldots,z_{2m})}=0$.

We use exactly the same process as to go from Prop.~\littlecancel\ to Thm.~\bigcancel.
We note that $\ket{\Psi(\ldots,z_i,\ldots,z_{i'},\ldots,z_{i''},\ldots)}$ is related to 
$\ket{\Psi(\ldots,z_i,z_{i'},z_{i''},\ldots)}$ by a product of $R$-matrices,
paying attention to the fact that none of these
$R$-matrices are singular for generic values of the other parameters (this is where we use
the fact that $i,i',i''$ are in cyclic order). Then we apply
Prop.~\littlecancelglob. \qed

\fig\basefig{Base link pattern of ${\cal L}_{3,8}$.}{\epsfxsize=3cm\epsfbox{base.eps}}
Let us now consider what we call the {\it base link pattern}\/ $\delta\in\LPl$ defined
by $\delta(i)=2n+1-i$, $1\le i\le 2n$, see Fig.~\basefig.
Theorem \bigcancel\ implies that
\eqn\base{
\Psi_{\delta}=\Omega \prod_{{\scriptstyle 1\le i<j\le m\atop{\rm or}}\atop\scriptstyle m+1\le i<j\le 2m}
\prod_{k=1}^{\ell} (q^k z_i-q^{-k}z_j)
}
where $\Omega$ is a polynomial to be determined. Thus,
$\deg\ket\Psi=\deg \Psi_{\delta}\ge \l m(m-1)$. 
Based on experience with similar models \refs{\DFZJ,\DFZJb,\DFZJc} in which one can prove a ``minimal
degree property'', as well as extensive computer investigations, it is reasonable to formulate the
\conj\degconj $\deg\ket\Psi=\l m(m-1)$.

One should be able to prove this conjecture either by {\it ad hoc}\/ methods, as in e.g.~\DFZJ,
or by a detailed analysis of the underlying representation theory on the space of polynomials,
as suggested by the work \Pas. 
This is not the purpose of the present work, and we proceed assuming Conjecture~\degconj.
To fix
the normalization of $\ket\Psi$ we set $\Omega=(-1)^{\ell m(m-1)/2}$ in Eq.~\base, so that the homogeneous value
$\Psi_\delta(1,\ldots,1)=\big({\ell+2\over2\sin(\pi/(\ell+2))}\big)^{m(m-1)}$ is positive.

\prop\partialdeg Assuming Conjecture~\degconj, each component $\Psi_\alpha$ is of degree at most $\ell(m-1)$
in each variable.

Proof. The proof is strictly identical to that of Thm.~4 of \DFZJ, and will be sketched only.
Using reflection covariance of the model, it is easy to see that
\eqn\refl{
\prod_{i=1}^{2m} z_i^d\ \Psi_{s\alpha}\left({1\over z_{2m}},\ldots,{1\over z_1}\right)
=\Psi_\alpha(z_1,\ldots,z_{2m})
}
where $s$ is the reflection of link patterns: $(s\alpha)(i)=2m+1-\alpha(2m+1-i)$,
and $d$ is the maximum degree of the components $\Psi_\alpha$ in each variable.
Equating the total degrees in all variables on both sides of Eq.~\refl, we find
$2md -\ell m(m-1)=\ell m (m-1)$, and therefore $d=\ell(m-1)$. \qed

We are now in a position to resolve the following natural question,
which is to ask what one can say about the {\it non-zero}\/ components when $z_j=q^{2k}z_i$.
Here we answer this question in the simplest situation:

\prop\recur Suppose $z_{i+1}=q^2 z_i$. Consider the embedding $\varphi_i$ of ${\cal L}_{\ell,2(m-1)}$ 
into $\LPl$ which inserts $2\ell$ sites at $S_i$, $S_{i+1}$ and $\ell$ arches between $S_i$ and $S_{i+1}$.
Then, assuming Conj.~\degconj,
\eqnn\recurpsi
$$\eqalignno{
\Psi_{\varphi_i(\alpha)}&(z_1,\ldots,z_{i+1}=q^2 z_i,\ldots,z_{2m})\cr
&=
q^{2(m-1)}
\prod_{j\ne i,i+1}\prod_{k=1}^\ell (z_i-q^{2k}z_j)
\Psi_\alpha(z_1,\ldots,z_{i-1},z_{i+2},\ldots,z_{2m})&\recurpsi
}$$
for all $\alpha\in{\cal L}_{\ell,2(m-1)}$,
where it is understood that on the r.h.s.\ $\Psi$ is the eigenvector at size $m-1$.

Proof. First we recall (cf proof of Prop.~\littlecancel) 
that $R_i(z_i,q^2 z_i)$ is proportional to $e_i^{(\ell)}$,
the projector onto the span of the image of $\varphi_i$, so that
according to Eq.~\YBE, $T(t|z_1,\ldots,z_i,z_{i+1}=q^2 z_i,\ldots,z_{2m})$ leaves 
this subspace invariant. This alone is sufficient to show that
$T(t|z_1,\ldots,z_i,z_{i+1}=q^2 z_i,\ldots,z_{2m})\varphi_i\propto\varphi_i
T(t|z_1,\ldots,z_{i-1},z_{i+2},\ldots,z_{2m})$, but
we need to compute the proportionality factor explicitly.
The latter is given by evaluating Fig.~\se.
Since the result is proportional to the projector $p$, one can close the outgoing lines,
replace the $R$-matrices with their expressions \defRb\ and then apply repeatedly Lemma~\eijb.
Simplifying Eq.~\defRb\ at $q=-\e{i\pi/(\ell+2)}$,
we find that the term $j,j'$ in the double sum produces a
contribution ${(z_i-t)(q\, z_i-q^{-1}t)\over (q^{-j-1}z_i-q^{j+1}t)(q^{-j}z_i-q^j t)}$
times the same for $j'$ with $z_i$ replaced with $z_{i+1}$ 
(noting in particular that the factors $1/(U_j U_{j'})$ produced by Lemma~\eijb\ 
compensate $a_ja_{j'}$). Finally we find that the coefficient of proportionality is
$\sum_{j=0}^\ell {(z_i-t)(q\, z_i-q^{-1}t)\over (q^{-j-1}z_i-q^{j+1}t)(q^{-j}z_i-q^j t)}=1$
times the same sum with $z_i$ replaced with $z_{i+1}$.
Thus,
\eqn\recurT{
T(t|z_1,\ldots,z_i,z_{i+1}=q^2 z_i,\ldots,z_{2m})\varphi_i=\varphi_i
T(t|z_1,\ldots,z_{i-1},z_{i+2},\ldots,z_{2m})
}
\fig\se{Contribution of $R(z_i,t)R(z_{i+1},t)$ to the sector where $S_i$ and $S_{i+1}$ are fully connected
to each other.}{\epsfxsize=8cm\epsfbox{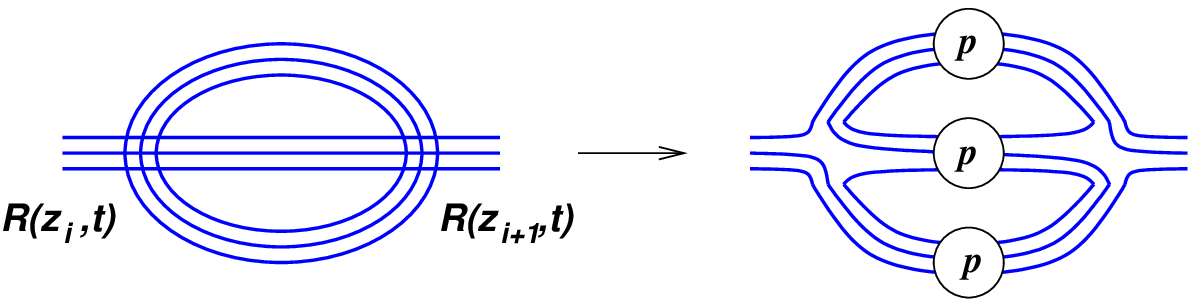}}
Lemma~\simpleev\ then implies that the l.h.s.\ and r.h.s.\ of Eq.~\recurpsi\ are proportional,
up to a rational function of the $z_i$ which is independent of $\alpha$. To fix the proportionality
factor, we consider the base link pattern, but rotated $i$ times in such a way that are $\ell$ arches
between $S_i$ and $S_{i+1}$.
When we remove the arches between $S_i$ and $S_{i+1}$ this is again the rotated base link pattern
but at size $2(m-1)$. We can therefore compare their expressions 
(Eq.~\base\ with $\Omega=(-1)^{\ell m(m-1)/2}$; this is the only place where we use Conj.~\degconj)
and collect the extra factors at
size $2m$. \qed

The theorem can be easily generalized to $z_j=q^2 z_i$, along the lines of Thm.~6 of \DFZJ, but this will
not be needed here. In the case $z_j=q^{2k} z_i$, $k>1$, the situation is more subtle: the recursion
would lead to a new type of ``mixed'' loop model with $2(m-1)$ usual subsets of $\ell$ vertices
and one special site which would have only $2(k-1)$ vertices fused together.
We do not pursue here this direction.

\example We provide the full analysis of the case $2m=4$.
As has already been mentioned in the course of the proof of Thm.~1, a state $\ket{j}$ in ${\cal L}_{\ell,4}$
is indexed by an integer $j$, $0\le j\le\ell$, in such a way that there are $\ell-j$ arches
between $S_1$ and $S_2$ and between $S_3$ and $S_4$, and $j$ arches between $S_2$ and $S_3$ and between
$S_4$ and $S_1$. (note that $\ket{\delta}=\ket{0'}=\ket\ell$). We immediately conclude from Prop.~\littlecancel\ 
that
\eqn\exfour{
\Psi_j(z_1,z_2,z_3,z_4)=\Omega_j \prod_{k=1}^j (q^{k} z_1-q^{-k} z_2)(q^k z_3-q^{-k} z_4)
\ \prod_{k=1}^{\ell-j} (q^k z_2-q^{-k} z_3)(q^k z_4-q^{-k} z_1)
}
where the $\Omega_j$ are constants if we assume the conjecture \degconj\ on the degree $2\ell$.
In order to determine them we consider the homogeneous situation i.e. the Hamiltonian
$H$. It is not too hard to compute off-diagonal elements of the matrix of $H$:
$H^j_{\ k}=2  U_{1+|j-k|}(\tau)$, $j\ne k$, and in particular to conclude that it is a symmetric matrix.
Therefore $\Psi_j$ must be proportional to $v_j=1/U_j(\tau)$. We compute
$\Psi_j(1,\ldots,1)=(-1)^\ell\big({\ell+2\over2\sin(\pi/(\ell+2))}\big)^2\,\Omega_j/U_j(\tau)^2$,
and using Eq.~\base\ to fix the normalization ($\Omega_\ell=\Omega=(-1)^\ell$), we find $\Omega_j=(-1)^\ell U_j(\tau)$.

Note that for $m>2$, Thm.~\littlecancelglob\ is not sufficient to determine up to a constant the entries $\Psi_\alpha$,
since they are in general not fully factorizable as products of $z_j-q^{2k}z_i$.

\subsec{Sum rule}
A very natural object is the pairing of the left eigenvector and of the right eigenvector:
we denote it by $Z(z_1,\ldots,z_{2m}):=\braket{0}{\Psi(z_1,\ldots,z_{2m})}$.

\prop\symsum $Z(z_1,\ldots,z_{2m})$ is a symmetric function of its arguments.

Proof. Start from $Z(z_1,\ldots,z_{i+1},z_i,\ldots,z_{2m})=\braket{0}{\Psi(z_1,\ldots,z_{i+1},z_i,\ldots,z_{2m})}$.
Applying Prop.~\exch, it is equal to $\bra{0}R_i(z_{i+1},z_i)\ket{\Psi(z_1,\ldots,z_{2m})}$.
On the other hand, from Lemma \explev, $\bra{0}R_i(z_{i+1},z_i)\ket{\cdot}=v\, R_i(z_{i+1},z_i)=v$. Thus,
$Z(z_1,\ldots,z_{i+1},z_i,\ldots,z_{2m})=Z(z_1,\ldots,z_{2m})$, which proves
the proposition. \qed

\thm\sumrule Assuming Conjecture \degconj,
$Z(z_1,\ldots,z_{2m})$ is the Schur function $s_{Y_{\ell,m}}(z_1,\ldots,z_{2m})$ 
associated to the Young diagram $((m-1)\ell,(m-1)\ell,\ldots,2\ell,2\ell,\ell,\ell)$.

Proof. In fact, we really claim 
that $s_{Y_{\ell,m}}(z_1,\ldots,z_{2m})$ is, up to multiplication by a scalar, the only symmetric 
polynomial of degree (at most) $\ell (m-1)$ in each variable,
which vanishes when the conditions of Thm.~\bigcancelglob\ are met. This clearly implies the
theorem (up to a multiplicative constant) due to Prop.~\symsum, Prop.~\partialdeg\ and Thm.~\bigcancelglob.

First, we show that $s_{Y_{\ell,m}}$ does satisfy these conditions. It is symmetric by definition, and its
degree in each variable is the width of its Young diagram that is $\ell(m-1)$. It can be expressed as
\eqn\xxx{
s_{Y_{\ell,m}}(z_1,\ldots,z_{2m})={\det_{1\le i,j\le 2m}(z_j^{h_i})\over \prod_{1\le i<j\le 2m} (z_i-z_j)}
}
where the $h_j$ are the shifted lengths of the rows of $Y_{\ell,m}$, that is
$h_{2i-1}=(i-1)(\ell+2) \equiv 0\ {\rm mod}\ \ell+2$, $h_{2i}=(i-1)(\ell+2)+1\equiv 1\ {\rm mod}\ \ell+2$, $i=1,\ldots,m$.
Assume now that $z_2=q^{2k}z_1$, $z_3=q^{2k'}z_2$. Isolate the three first columns of the determinant in the
numerator of Eq.~\xxx:
the odd rows are of the form $z_1^{h_{2i-1}} (1,1,1)$ whereas the even rows are of the form
$z_1^{h_{2i}} (1,q^{2k},q^{2(k+k')})$. Thus, we have two series of $m$ proportional rows: this proves that
the $3\times 2m$ matrix is of rank 2, and that the full $2m\times 2m$ matrix is singular.
If the $z$'s are distinct the denominator of Eq.~\xxx\ is non-zero and we conclude that $s_{Y_{\ell,m}}$ vanishes.

Next, we show that it is the only such polynomial by induction. The step $m=0$ is trivial.

At step $m$, 
consider a symmetric polynomial $Z(z_1,\ldots,z_{2m})$ in $2m$ variables, of degree $\ell (m-1)$ in each variable,
which vanishes when the conditions of Thm.~\bigcancelglob\ are met. 
Note that since $Z$ is symmetric, the conditions
can be in fact extended to arbitrary distinct integers $(i,i',i'')$.
Setting $z_j=q^{2k} z_i$, $1\le k\le \ell+1$, $i\ne j$, they therefore imply the following factorization:
\eqn\vani{
Z|_{z_j=q^{2k}z_i}=
\Bigg(
\prod_{h\ne i,j}\prod_{\scriptstyle p=1\atop\scriptstyle p\ne k}^{\ell+1} (z_h-q^{2p} z_i)
\Bigg)
\,
W(z_1,\ldots,\hat{z}_i,\ldots,\hat{z}_j,\ldots,z_{2m})
}
where $W$ is a symmetric polynomial of the $2m-2$ variables $z_h$, $h\ne i,j$, 
of degree $\ell (m-2)$ in each, which still
vanishes when the conditions of Thm.~\bigcancelglob\ are met. $W$ does not depend
on $z_i$ because the $2\ell(m-1)$ prefactors exhaust the degree of $z_i$ and $z_j$ combined.
The induction
hypothesis implies that $W={\rm const}\ s_{Y_{\ell,m-1}}(z_1,\ldots,\hat{z}_i,\ldots,\hat{z}_j,\ldots,z_{2m})$.
The constant is independent of $i$ or $j$ by symmetry; and of $k$, as one can check by taking $z_i\to\infty$ 
(indeed in the limit $z_i$, $z_j\to\infty$, $Z$ must be proportional to $(z_i z_j)^{\ell(m-1)}$, which fixes the
relative normalization of $Z|_{z_j=q^{2k}z_i}$ for varying $k$).
$Z$, as a function of a given $z_j$, is thus specified at $(\ell+1) (2m-1)$ points by
Eq.~\vani; this is enough to determine uniquely a polynomial of degree $\ell(m-1)$.
Therefore $Z={\rm const}\ s_{Y_{\ell,m}}(z_1,\ldots,z_{2m})$, which
concludes the induction.

Finally, one fixes the constant by another induction using Prop.~\recur\ 
(Eq.~\recurpsi). \qed

Note the obvious

\corol $\braket{\Psi}{\Psi}=s_{Y_{\ell,m}}^2$.

A final remark concerns the homogeneous situation where all $z_i$ are equal. 
In this case one can evaluate explicitly
the Schur function i.e.\ the dimension of the corresponding $sl(2m)$ representation:
\eqn\homo{
Z(1,\ldots,1)=((\ell+2)i)^{m(m-1)} \prod_{i,j=1}^m {(\ell+2)(j-i)+1\over j-i+m}\ .
}

\table\thetable{First few values of $Z(1,\ldots,1)$.}{
\vbox{\offinterlineskip\halign{\strut\hfil$#$\quad&\vrule#&&\ \hfil$#$\hfil\crcr
\vtop{\hbox{\ \rlap{$m$}}\vskip2pt\hbox{$\ell$}}&& 1 & 2 & 3 & 4 & 5&6\cr
\omit&height2pt\cr
\noalign{\hrule}
\omit&height2pt\cr
1&& 1& 6& 189& 30618& 25332021& 106698472452\cr
2&& 1& 20& 6720& 36900864& 3280676585472& 4702058148658151424\cr
3&& 1& 50& 103125& 8507812500& 27783325195312500&
     3574209022521972656250000\cr
4&& 1& 105& 945945& 707814508401& 43505367274327463505&
     218541150429748620278689395225\cr
5&& 1& 196& 6117748& 29406803321896&
     21520945685492367246132& 2385377935975138162776292257847164\cr
}
}}

\newsec{Conclusion}
This paper has tried to demonstrate the power of the methods devised
in \DFZJ\ and subsequent papers by applying it to the case of fused 
$A_1$ models. A special point has been found for each such model
-- which is nothing but the point at which the central charge of the infrared fixed point vanishes.
We call this point ``combinatorial'' because one can hope that the
properties it possesses have interesting combinatorial meaning. Some of it have been
described in the paper: existence of a left eigenvector with a simple form
in the basis that we have built; simple sum rule. However, many questions remain
open.

First and foremost, one would like to have a generalized Razumov--Stroganov \RS\ conjecture for these
fused models. In the present case, it would correspond to identifying each component of the ground
state eigenvector of the Hamiltonian with the $\tau$-enumeration of some combinatorial objects. By $\tau$-enumeration
we mean that the enumeration should be somehow weighted with $\tau$ to take into account the fact that
the components belong to ${\Bbb Z}[\tau]$ (in the unfused case they are integers). For example, note that
at $\ell=2$ we do know an interpretation of the sum of all components: up to a missing factor $2^m$
(which can be naturally introduced in the normalisation of $v$), it is the $2$-enumeration of Quarter-Turn Symmetric 
Alterating Sign Matrices (QTSASM) \refs{\Kup,\Oka}. The introduction of spectral parameters and the appearance of the
Schur function of Thm.~\sumrule\ also arise in this context.
One should explore how this connection can be extended at the level
of each component.
Note that in the ASM literature, $1-$, $2-$ and $3-$enumerations are often considered.
In our language, these are really $1-$, $\sqrt{2}-$, $\sqrt{3}-$enumerations, which
correspond to $\ell=1,2,4$.

Also, many additional properties should be obtainable, along the lines of the abundant literature on the
unfused case. For example we propose here the following
\conj\largest Denote $\Psi(m)\equiv \Psi(z_1=\cdots=z_{2m}=1)$, $Z(m)\equiv
Z(z_1=\cdots=z_{2m}=1)$. Then $\Psi_0(m)$ is the largest entry of $\ket{\Psi(m)}$ (where we recall
that $0$ is the pattern that fully connects $S_{2i-1}$ and $S_{2i}$), and
$${\Psi_0(m)\over\Psi_\delta(m)}={Z(m-1)\over\Psi_\delta(m-1)}$$
In other words, if $\Psi$ is normalized in such a way that the base link pattern has entry 1, the largest
entry at size $m$ is the (weighted) sum at size $m-1$.

Equally interesting is the study of the space of polynomials spanned by the components of the ground state
eigenvector and the related representation theory, following the philosophy of \Pas. One should emphasize the
difficulty of such a task, because it involves separating the action on polynomials from the action on
link patterns -- even in the case of the Birman--Wenzel--Murakami algebra (BWM) this difficulty appears
\refs{\DFZJb,\Pasb}, and for us BWM is only the simplest fused case (corresponding to $\ell=2$).

Closely related is the extension of this work to a generic value of $q$ by introducing an approprate
quantum Knizhnik--Zamolodchikov ($q$KZ) equation. Clearly all arguments of Sect.~3.3 depend only
on polynomiality of the ground state eigenvector and on Prop.~\exch,
which is the key equation of the $q$KZ system.
A natural conjecture is that at $q$ generic will 
appear precisely the $q$KZ equation at level $\ell$, which would be 
of greater interest than the level $1$
(``free boson'') $q$KZ of the unfused model.

One should also be able to combine the ideas of \DFZJc\ and of the present work to study fused
higher rank models; it is easy to guess the kind of properties they will possess at the point
$q=-\e{\pm i\pi/(k+\ell)}$, $k$ dual Coxeter number. One could also consider fused models
with other boundary conditions (open boundary conditions, etc, as in \refs{\DFb,\DFZJd}).


Finally, it would be interesting to find 
some relation between our formulae, and in particular the sum rule,
with the recent work \Kit\ which generalizes the domain wall boundary conditions
of the six-vertex model (relevant to the sum rule of the unfused loop model) to fused models.

\centerline{\bf Acknowledgments}
The author thanks P.~Di Francesco and J.-B.~Zuber for numerous discussions,
and acknowledges the support
of European networks ``ENIGMA'' MRT-CT-2004-5652, ``ENRAGE'' MRTN-CT-2004-005616,
and of ANR program ``GIMP'' ANR-05-BLAN-0029-01.

\appendix{A}{Projection operator}
Following \Ma\ (see also \KS\ and references therein), 
we define recurrently the projectors $p^{(k)}$ by $p^{(1)}=1$ and
\eqn\defproja{
p^{(k+1)}(e_j,\ldots,e_{j+k-1})
=p^{(k)}(e_j,\ldots,e_{j+k-2})(1-\mu_k(\tau) e_{j+k-1})
p^{(k)}(e_j,\ldots,e_{j+k-2})
}
where $\mu_k(\tau)=U_{k-1}(\tau)/U_k(\tau)$ and $U_k$ is the Chebyshev
polynomial of the second kind. The projectors used in this paper are simply
$p_i:=p^{(\ell)}(e_{\ell (i-1)+1},\ldots,e_{\ell i-1})$.

By recursion on $k$ one can prove the following facts:

\item{(a)} $(p^{(k)}(e_j,\ldots,e_{j+k-2})e_{j+k-1})^2=\mu_k^{-1}
p^{(k)}(e_j,\ldots,e_{j+k-2})e_{j+k-1}$ and $p^{(k)}{}^2=p^{(k)}$.
\item{(b)} $p^{(k)}$ is $\star$-symmetric.
\item{(c)} $p^{(k)}(e_j,\ldots,e_{j+k-2}) 
e_m=e_m p^{(k)}(e_j,\ldots,e_{j+k-2})=0$
for $m=j,\ldots,j+k-2$. (use (a) for $m=j+k-2$).
\item{(d)} $p^{(k)}$ is ``left-right symmetric'', that is it also satisfies
\eqn\defprojb{
p^{(k+1)}(e_j,\ldots,e_{j+k-1})
=p^{(k)}(e_{j+1},\ldots,e_{j+k-1})(1-\mu_k(\tau) e_j)
p^{(k)}(e_{j+1},\ldots,e_{j+k-1})
}
\item{(e)} $p^{(k)}p^{(k')}=p^{(k)}p^{(k')}=p^{(k)}$ when $k\ge k'$ and the arguments of $p^{(k')}$ are
a subset of those of $p^{(k)}$ (use Eqs.~\defproja\ and \defprojb).

Note that property (i) of Sect.~(2.3) is a direct consequence of (c).

\appendix{B}{Enumeration of admissible states}
Call $W_{\ell,m}$ the set of Lukacievicz words of length $(\ell+1)m$
taking values in $\{ \ell,-1\}$, that is
\eqn\defwords{
W_{\ell,m}=\left\{ w\in \{\ell,-1\}^{(\ell+1)m} : \sum_{i=1}^j w_i \ge 0\ \forall j<(\ell+1)m,
\ \sum_{i=1}^{(\ell+1)m}w_i=0\right\}
}
These words describe rooted planar trees with arity $\ell+1$, and it is well-known that
\eqn\cardwords{
\# W_{\ell,m}={((\ell+1)m)!\over (\l m+1)!m!}
}
We shall therefore describe a bijection between $\LPl'$ and $W_{\ell,m}$.

Start from a link pattern $\alpha$. As an intermediate step it is convenient to rewrite it as a
Dyck word $w$ 
(the case $\ell=1$ of the Lukacievicz words above). Considering the link pattern as unfolded
in the half-plane, we associate to each vertex where an arch starts (resp.\ ends) a $+1$ (resp.\ $-1$).
This is in fact the bijection in the case $\ell=1$. We shall now restrict ourselves to $\ell$-admissible
link patterns. The goal is to transform the word $w$
by condensing groups of $\ell$ ``$+1$'' into a single ``$\ell$''.

We read the word $w$ from left to right, in sequences of $\ell$ letters. Since $\alpha\in\LPl$,
these sequences can only be $k$ ``$-1$'' followed by $\ell-k$ ``$+1$'', $0\le k\le\ell$. We distinguish
three cases:

\item{(i)} $k=0$: if there are only ``$+1$'', replace them with a single ``$\ell$''.

\item{(ii)} $k=\ell$: if there are only ``$-1$'', leave them intact.

\item{(iii)} $0<k<\ell$: the $k$ ``$-1$'' are left intact.
As to the $\ell-k$ ``$+1$'', two situations arise. Either (iiia) they have not been flagged yet, in which
case they are replaced with a single ``$\ell$''. Say the first $+1$ of the sequence is at position $i$.
Find the first position $j$ for which $\sum_{p=i+1}^{j-1} w_p<0$. According to Lemma~\parity,
we know that $r(i)+r(j-1)=\ell-1$, and that $w_j$ and all its successors are $+1$
(there are $\ell-r(i)=\ell-k$ of them). We flag them.
Or (iiib) they have been flagged, in which case we ignore them.

It is easy to show that the resulting word is indeed in $W_{\ell,m}$. In particular,
the $\ell$-admissibility ensures that sequences with $k$ ``$+1$'' with $0<k<\ell$ always come in pairs,
the second one being flagged.

Inversely, start from a word $w\in W_{\ell,m}$. Read it from left to right.
Each time we come across a ``$\ell$'' at position $i$ (all modifications to the left being taken into
account in the position), with $r(i)=k$, we replace it using the following rule:

\item{(i)} $k=0$: we simply replace it with a sequence of $\ell$ ``$+1$''.

\item{(ii)} $k>0$: we replace it with $\ell-k$ ``$+1$''. Then we look for the first position $j$
such that $\sum_{p=i+1}^{j-1} w_p<0$ (being careful that the sum starts with $\ell-k-1$ newly
created ``$+1$'').
We insert $k$ extra ``$+1$'' between positions $j$ and $j+1$.

This will clearly produce a Dyck word, and it not hard to check that the corresponding link
pattern is $\ell$-admissible. The two operations described above being clearly inverse of each other,
we conclude that they are bijections.

\example
these are the words associated to ${\cal L}'_{2,6}$,
with the same ordering as in Sect.~2.5:
$$W_{2,3}=\left\{\matrix{
 2 & -1 & -1 & 2 & -1 & -1 & 2 & -1 & -1 \cr
 2 & -1 & -1 & 2 & -1 & 2 & -1 & -1 & -1 \cr
 2 & -1 & -1 & 2 & 2 & -1 & -1 & -1 & -1 \cr
 2 & -1 & 2 & -1 & -1 & -1 & 2 & -1 & -1 \cr
 2 & 2 & -1 & -1 & -1 & -1 & 2 & -1 & -1 \cr
 2 & -1 & 2 & -1 & -1 & 2 & -1 & -1 & -1 \cr
 2 & -1 & 2 & -1 & 2 & -1 & -1 & -1 & -1 \cr
 2 & -1 & 2 & 2 & -1 & -1 & -1 & -1 & -1 \cr
 2 & 2 & -1 & -1 & -1 & 2 & -1 & -1 & -1 \cr
 2 & 2 & -1 & -1 & 2 & -1 & -1 & -1 & -1 \cr
 2 & 2 & -1 & 2 & -1 & -1 & -1 & -1 & -1 \cr
 2 & 2 & 2 & -1 & -1 & -1 & -1 & -1 & -1
}\right\}$$

\listrefs
\end